\documentclass[aps,preprint,showpacs,groupedaddress]{revtex4}
\usepackage {graphicx,epsfig,graphics,color}

\begin{document}

\title{Spatial characterization of the magnetic field profile
of a probe tip used in magnetic resonance force microscopy}

\author{E. Nazaretski}
\affiliation{Los Alamos National Laboratory, Los Alamos, NM
87545}
\author{E. A. Akhadov}
\affiliation{Los Alamos National Laboratory, Los Alamos, NM 87545}
\author{I. Martin}
\affiliation{Los Alamos National Laboratory, Los Alamos, NM 87545}
\author{D. V. Pelekhov}
\affiliation{Department of Physics, Ohio State University, Columbus
OH 43210}
\author{P. C. Hammel}
\affiliation{Department of Physics, Ohio State University, Columbus
OH 43210}
\author{R. Movshovich}
\affiliation{Los Alamos National Laboratory, Los Alamos, NM
87545}

%\date{\today}

\begin{abstract}
We have developed the experimental approach to characterize spatial
distribution of the magnetic field produced by cantilever tips used
in magnetic resonance force microscopy (MRFM). We performed MRFM
measurements on a well characterized diphenyl-picrylhydrazyl (DPPH)
film and mapped the 3D field profile produced by a {\em
$Nd_2Fe_{14}B$} probe tip. Using our technique field profiles of
arbitrarily shaped probe magnets can be imaged.
\end{abstract}

%\pacs{07.79.Pk, 07.55.-w, 76.50.+g, 75.70.-i}
\maketitle

%\section{INTRODUCTION}
Magnetic resonance force microscopy attracted a lot of interest in
the last few years due to its high force sensitivity and excellent
spatial resolution of magnetic properties. MRFM has been used in
studies of electron and nuclear spin systems culminating in the
detection of the force signal originating from a single electron
spin \cite{Rugar 2004}. Recent experiments on nuclear spins of
$^{19}F$ in $CaF_2$ samples demonstrated the spatial resolution of
90 nm \cite{Mamin 2007}, orders of magnitude better than
conventional magnetic resonance imaging technique. In the long term,
MRFM is envisioned as a possible route to achieve imaging of
individual molecules. Experiments on ferromagnetic systems showed
the potential for spatially resolved ferromagnetic resonance in
continuous and microfabricated samples \cite{Nazaretski 2007, Mewes
2006}. In MRFM experiments, force {\bf F} exerted on a cantilever,
is a convolution of the sample's magnetization and the gradient of
the magnetic field produced by the probe tip. To perform correct
imaging, quantitative knowledge of the {\em spatial} distribution of
the tip field is required. At present, the most common way to
characterize magnetic tips is to use the cantilever magnetometry
\cite{Rossel 1996, Stipe 2001}. It provides information about the
magnetic moment of the tip {\bf m}, however, it is also sensitive to
the relative orientation of {\bf m} with respect to the external
magnetic field and the direction of cantilever's oscillations.
Moreover, the detailed spatial field profile of the magnetic tip can
not be inferred. Alternative approach utilizes the spectroscopic
nature of MRFM and has been demonstrated in previous studies
\cite{Mamin 2007,Chao 2004,Wago 1998, Bruland 1998, Hammel 2003}. In
these experiments the strength of the probe field has been
determined from the position of the onset in the MRFM spectra as a
function of the probe-sample separation $z$. Based on this
information, the point dipole approximation has been used to model
the magnetic tip. The situation becomes more complicated if the
shape of the tip is irregular or {\bf m} is tilted with respect to
the $\hat{z}$ direction. Under these circumstances the
one-dimensional approach is insufficient, and does not reveal the
spatial field profile of the probe tip. In this letter we propose a
method for detailed mapping of the tip magnetic
field, free of any assumptions about the tip shape, size, or composition.\\
In MRFM experiments the magnetic tip of a cantilever is used to
generate the inhomogeneous magnetic field causing {\em local}
excitation of the spin resonance in a small volume of the sample
known as {\em sensitive slice}. The resonance condition is written
as follows
\begin{equation}
|{\bf H}_{tot}(r)|=\frac{\omega_{RF}}{\gamma}, \label{Equation 1}
\end{equation}
where $\gamma$ is the gyromagnetic ratio. The total field ${\bf
H}_{tot}(r)$ can be expressed as
\begin{equation}
{\bf H}_{tot}(r)={\bf H}_{ext}+ {\bf H}_{tip}(r), \label{Equation 2}
\end{equation}
where ${\bf H}_{ext}$ is the externally applied magnetic field and
${\bf H}_{tip}(r)$ is the field of the probe tip.
%${\bf H}_{tip}(r)$ also provides coupling of the sample's
%magnetization ${\bf M}(r,t)$ to the cantilever.
Width $\Delta z$ of the sensitive slice is determined by the ratio
of the resonance linewidth $\Delta H_{res}$ and the strength of the
gradient field $\nabla H_{tip}$ produced by the probe tip, $\Delta
z$ = $\frac{\Delta H}{|\nabla H_{tip}|}$ \cite{Suter 2002}.  Three
dimensional images of electron spin densities can be reconstructed
by performing lateral and vertical scanning of the sensitive slice
across the sample\cite{Wago 1998,Chao 2004}.

\begin{figure}[h]
\includegraphics [angle=0,width=5.5cm]{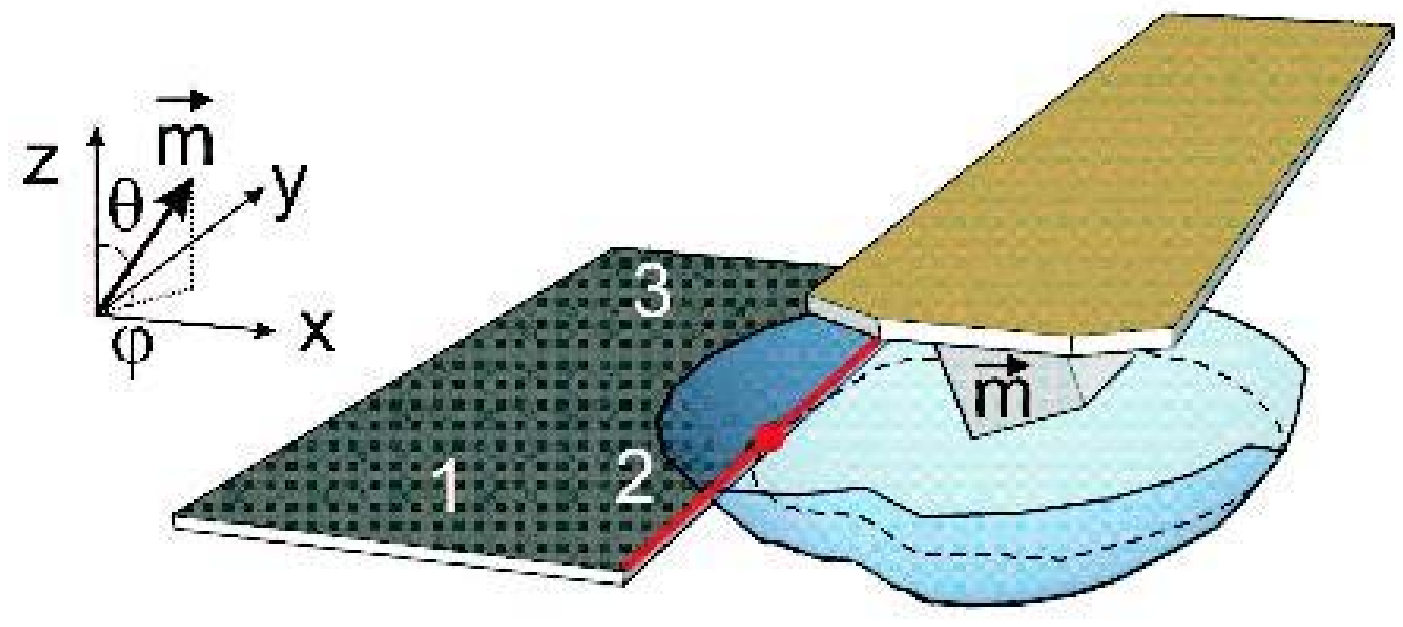}
\caption{} \label{Figure 3}
\end{figure}

The concept behind our method for detailed characterization of the
tip field profile is illustrated in Fig.~\ref{Figure 3}.  It
requires a thin-film sample with sharp edges. When the sensitive
slice {\em touches} the sample edge, a {\em leading edge} signal is
detected.  At this location, the sample edge is a tangent line to
the sensitive slice for a reasonable magnetic tip. Thus, scanning in
3D and recording the locations corresponding to the leading edge
enables full reconstruction of the sensitive slice.  If desired, it
can be then parameterized using dipolar, quadrupolar, etc moments.

\begin{figure}[h]
\includegraphics [angle=0,width=8.5cm]{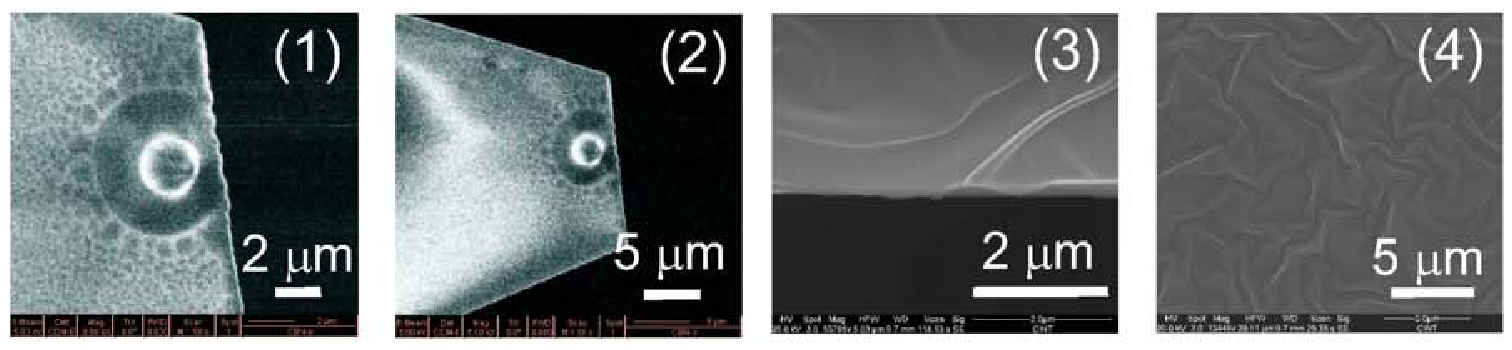}
\caption{} \label{Figure 1}
\end{figure}

To illustrate this procedure, we report on MRFM measurements on a
well characterized DPPH film, while laterally scanning the
cantilever over its edge.  We used a commercially available Veeco
$Si_3N_4$ cantilever with the resonance frequency of $\approx$ 8 kHz
and the spring constant $k$ of $\approx$ 0.01 N/m \cite{Veeco}. The
original tip was removed by focused ion milling and a small magnetic
particle of $Nd_2Fe_{14}B$ available from Magnequench Inc.
\cite{Magnequench} has been glued to the end of a cantilever with
Stycast 1266 epoxy in the presence of an aligning magnetic field.
Consequently, the tip has been magnetized in the field of 80 kOe.
The MRFM tip has a spherical shape with the diameter of $\approx$
2.4 $\mu$m and its SEM images are shown in panels (1) and (2) in
Fig. \ref{Figure 1}. The saturation magnetization of {\em
$Nd_2Fe_{14}B$} particles has been measured in a SQUID magnetometer,
and is equal to $4\pi M_s$ = 13 kG \cite{Nazaretski 2006a}. Based on
the SEM image we estimate the probe moment to be
(7.5$\pm$0.4)$\times$10$^{-9}$ emu, in agreement with the value of
(6.9$\pm$0.5)$\times$10$^{-9}$ emu measured by the cantilever
magnetometry. The cantilever is mounted on top of a double scanning
stage of a low temperature MRFM system \cite {Nazaretski 2006,
Attocube} . For data acquisition, the temperature was stabilized at
10 K and the amplitude modulation scheme has been implemented to
couple to the in-resonance spins. The DPPH powder \cite {DPPH} was
dissolved in acetone and deposited on a 100 $\mu$m thick silicon
wafer in a spin-coater at 3000 rpm. To protect the film, 20 nm of Ti
was deposited on top of DPPH. Approximately 2$\times$1.6 mm$^2$
piece was cleaved from a wafer and glued to the strip-line resonator
of the microscope. The structure of the film and sharpness of edges
were inspected in SEM and are shown in Fig. \ref{Figure 1}. The film
was found to be continuous, and its thickness varied between 400 and
600 nm.

\begin{figure}[h]
\includegraphics [angle=0,width=8.5cm]{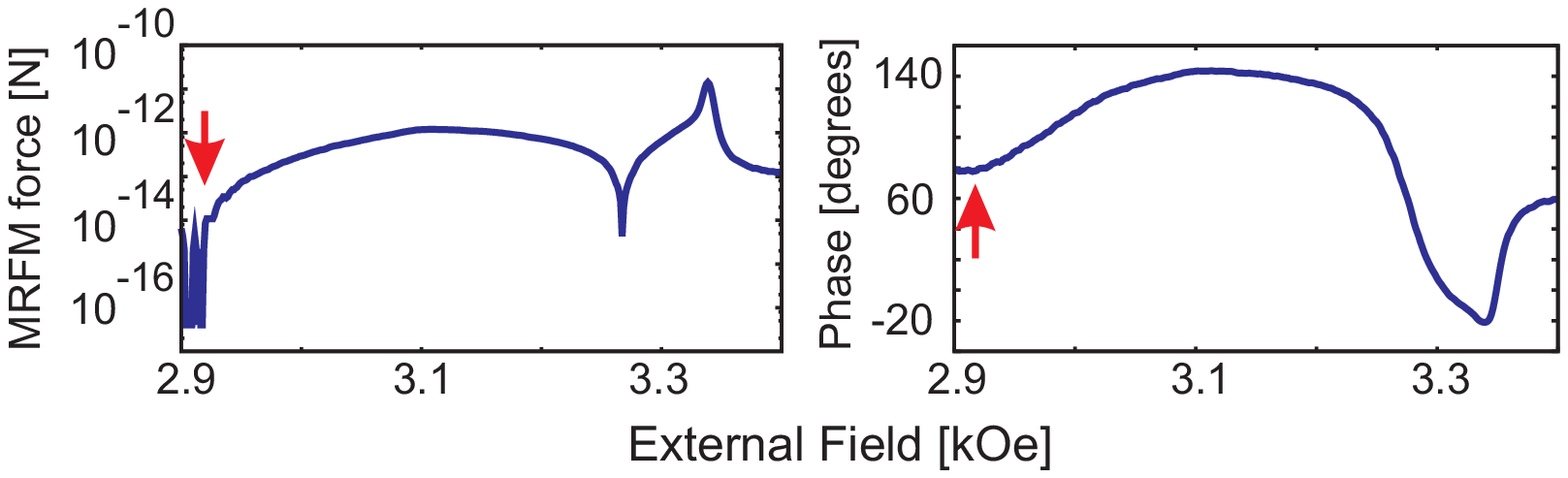}
\caption{} \label{Figure 2}
\end{figure}

Fig. \ref{Figure 2} shows the typical MRFM spectrum recorded in a
DPPH film. When the tip is located above the film, the strongest tip
field experienced by the sample is situated directly under the probe
magnet (assuming ${\bf m}$ $\parallel$ ${\bf H}_{ext}$). The field
value in the MRFM spectrum where the sensitive slice just touches
the DPPH film is called the {\em leading edge} \cite{Suter 2002},
and is indicated by arrows in Fig. \ref{Figure 2}.

The large positive peak at $\approx$ 3.34 kOe corresponds to the
bulk-like resonance. It originates from the large region of the
sample where the tip field is small, but due to the large number of
spins the MRFM signal is significant. The field difference between
the bulk-like resonance and the position of the leading edge
provides the {\em direct} measure of the probe field strength.

\begin{figure}[h]
\includegraphics [angle=0,width=8.5cm]{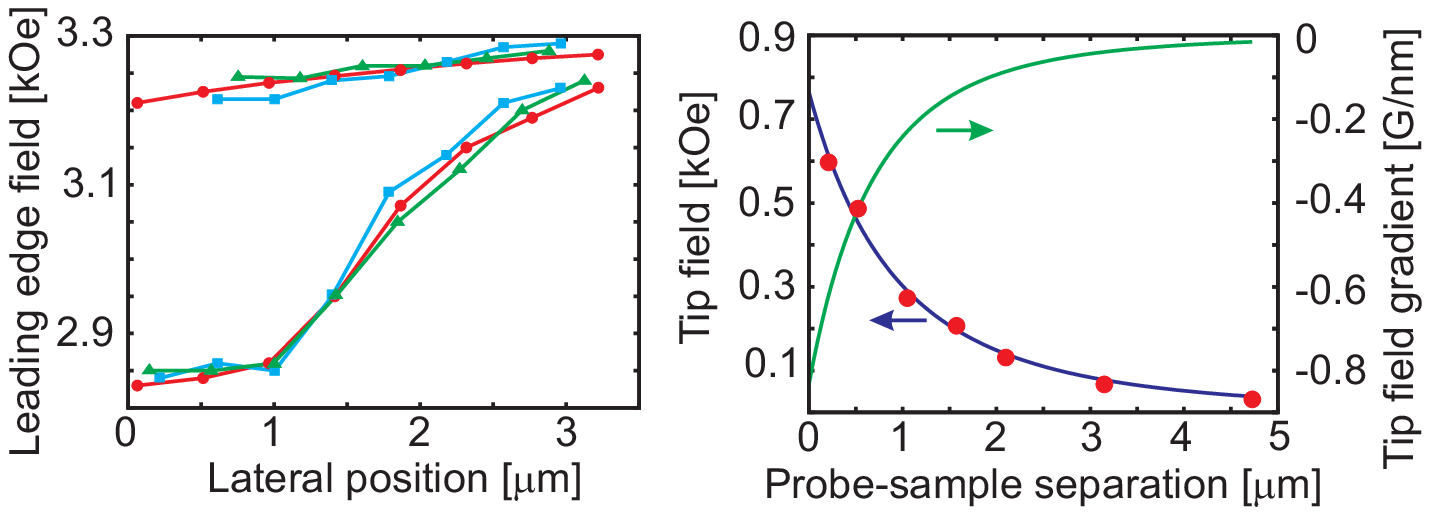}
\caption{} \label{Figure 4}
\end{figure}

\begin{figure}[h]
\includegraphics [angle=0,width=8.5cm]{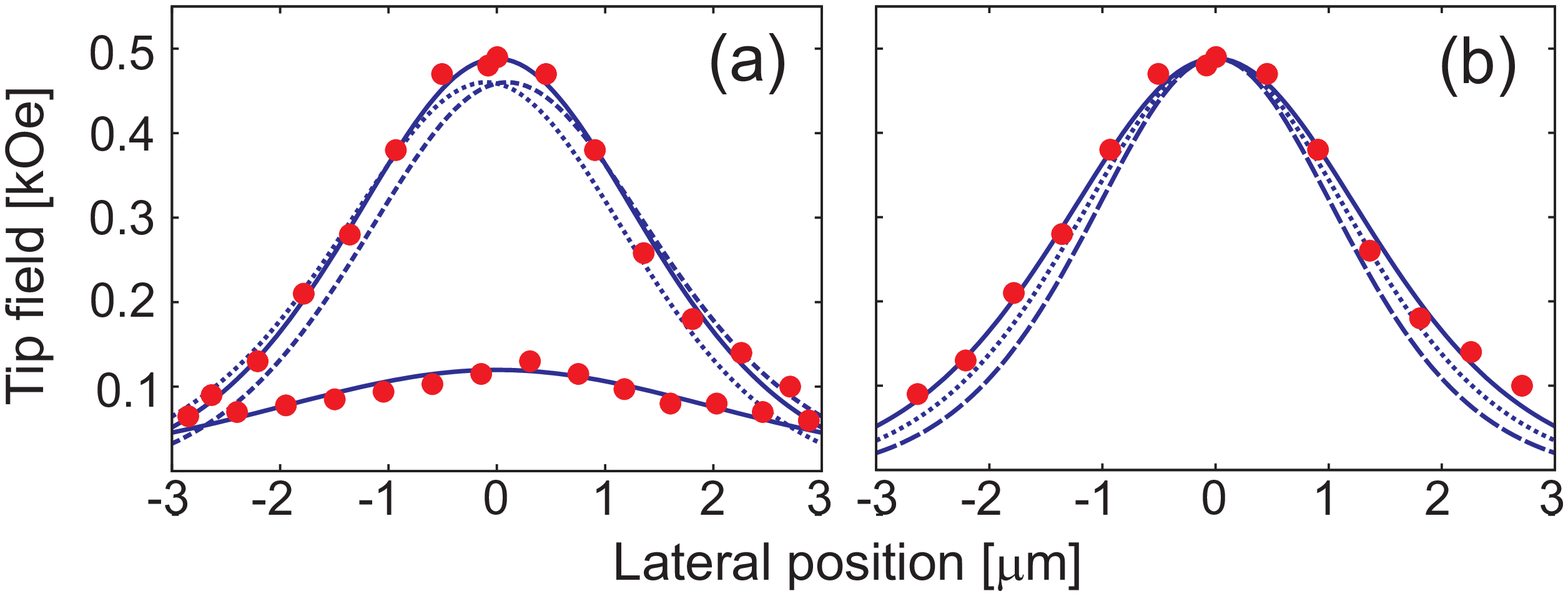}
\caption{} \label{Figure 6}
\end{figure}

Fig. \ref{Figure 3} shows the schematic of the characterization
experiment. We fixed the probe-sample separation $z$, and approached
different edges of the DPPH film while tracking the leading edge.
The left panel of Fig. \ref{Figure 4} shows the field evolution of
the leading edge for two values of $z$ and three different
directions of lateral scanning over the film edge. The almost
identical shape of the curves indicates that {\bf m} is
approximately parallel to the direction of ${\bf H}_{ext}$. In the
first approximation, our tip can be modeled as a magnetic dipole.
The field profile produced on the surface of the sample can be
written as follows \cite{Jackson 1975}:

\begin{eqnarray}
H(R, \theta, \varphi)& = \frac{4\pi M_s r_0^3}{3}\times \{
\frac{-3z(\sin\theta(x\sin\varphi+y\cos\varphi))}{R^{5}}+\nonumber \\
& +\frac{3z^2\cos\theta}{R^{5}}-\frac{\cos\theta}{R^{3}}\}, \label{Equation 3}
\end{eqnarray}
%\begin{equation}
%H_z(x)=\frac{4\pi M_s r_0^3}{3}\frac{(2z^2-x^2)}{(z^2+x^2)^{5/2}},
%\label{Equation 3}
%\end{equation}

\noindent where $4\pi M_s$ is the saturation magnetization of
$Nd_2Fe_{14}B$, $r_0$ is the radius of the tip, $R$ is the vector to
the point where the field is determined, %$R$ can also be written as
%$R = \sqrt{x^2+y^2+z^2}$.
$\theta$ and $\varphi$ are the angles which describe the spatial
orientation of {\bf m} (see Fig. \ref{Figure 3}). \\The right panel
of Fig. \ref{Figure 4} shows the z-component of the probe field on
the sample's surface as a function of $z$. Solid line is the fit
using Eq. \ref{Equation 3} and assuming parallel orientation of {\bf
m} and ${\bf H}_{ext}$. Fig. \ref{Figure 6}(a) shows the comparison
between the lateral field profile of the tip simulated according to
Eq. \ref{Equation 3}, and the actual data points taken from the left
panel of Fig. \ref{Figure 4}. Good agreement between the observed
and expected behavior suggests that, indeed, our probe tip can be
approximated as a dipole, and its magnetization is aligned along the
direction of ${\bf H}_{ext}$.
% within  $\pm$ 7$^\circ$.
In case of any significant misalignment the tip field profile would
change substantially, as shown in Fig. \ref{Figure 6}(a). For both
simulations shown in Fig. \ref{Figure 4} and \ref{Figure 6}, we had
to offset the probe-sample separation by 1.42 $\pm$ 0.03 $\mu$m ($z$
is the only free parameter in the fit) which suggests that due to
the short range probe-sample interaction the cantilever snaps to the
sample at distances smaller than 1.42 $\mu$m \cite{Berger 1999,
Dorofeyev 1999}. The presence of an offset may indicate the reduced
magnetic moment of the tip. However, our cantilever magnetometry
measurements of the tip moment agree well with the expected value,
as mentioned earlier in the paper. Moreover, in Fig. \ref{Figure
6}(b) we show the calculated spatial field profile of 2 $\mu$m, 2.2
$\mu$m and 2.4 $\mu$m diameter tips. The fit for the 2.4 $\mu$m
diameter tip provides the best agreement with the data points.
Another argument in support of our tip model pertains to the
magnitude of the MRFM force exerted on a cantilever in a particular
sensitive slice. In Fig. \ref{Figure 2} we take the measured MRFM
force at $H_{ext}$ = 3.038 kOe and compare it to our estimates. The
calculations yield the force value of $\approx$
6.9$\times$10$^{-13}$ N in good agreement with the measured value of
5.7$\times$10$^{-13}$ N. Thus, dipolar approximation and our
assumptions for the tip moment were adequate for the present
experiment. Importantly, the same technique could be applied to map
field profile from a more irregular tip.

In summary, we have studied the evolution of locally excited
electron-spin resonance in a DPPH film. By tracking the position of
the leading edge in MRFM spectra for different hight and direction
of the approach to the sample, we have determined the spatial field
profile of the cantilever tip. Measuring the MRFM signal onset over
the large range of positions with adequate sensitivity allows to
deconvolve the spatial field profile produced by arbitrarily shaped
magnetic tips used in the magnetic resonance force microscopy.

This work was supported by the US Department of Energy and was
performed, in part, at the Center for Integrated Nanotechnologies at
Los Alamos and Sandia National Laboratories. Personnel at Ohio State
University was supported by the US Department of Energy through
grant DE-FG02-03ER46054.

\newpage

Figure Caption

\vspace{0.5cm}

FIG.1 Schematic of the tip characterization technique. Detection of
the {\em leading edge} signal indicates that the sample edge is {\em
tangent} to the sensitive slice. 3D scanning can thus be used to
fully reconstruct the shape of the sensitive slice.

\vspace{0.5cm}

FIG.2 Panel (1)and (2): SEM images of the probe magnet. Panel (3)
shows the edge of the DPPH film and panel (4) is the top view
showing fine structures on the surface of the film.

\vspace{0.5cm}

FIG.3 Amplitude and phase of the MRFM signal recorded at $T$ = 10 K,
$\omega_{RF}$ = 9.35 GHz, $z$ = 0.73 $\mu$m. The position of the
{\em leading edge} is indicated by arrows.

\vspace{0.5cm}

FIG.4 Left panel: field evolution of the leading edge as a function
of lateral position over the DPPH film edge. The upper and lower set
of curves correspond to $z$ = 2.35 $\mu$m and $z$ = 0.53 $\mu$m
respectively. Circles represent the approach of the sample from side
'1', squares from side '2' and triangles form side '3' of the sample
as shown in Fig. \ref{Figure 3}. Right panel:  the $z$-component of
the tip field as a function of the probe-sample separation (left
Y-axis) and the corresponding field gradient (right Y-axis). Solid
curve is the fit to Eq. \ref{Equation 3}.

\vspace{0.5cm}

FIG.5 (a) Lateral field profile of the tip for approaches of sides
'1' and '3' of the sample, as shown in Fig. \ref{Figure 3}. Data
points are taken from the left panel in Fig. \ref{Figure 4}. '0' on
the X-axis corresponds to the edge of the film. Upper and lower data
points correspond to $z$ = 0.53 $\mu$m and $z$ = 2.35 $\mu$m
respectively. Solid curve is fitted to the data using Eq.
\ref{Equation 3}. Dotted and dashed lines show the expected field
profile of the tip where $\theta$ = $\varphi$ = 20$^\circ$ and
$\theta$ = -20$^\circ$, $\varphi$ = 20$^\circ$ respectively. (b)
expected field profile for the tip with $r_0$=1.2 $\mu$m,
z-offset=1.4 $\mu$m (solid line), $r_0$=1.1 $\mu$m, z-offset=1.12
$\mu$m (dotted line) and $r_0$=1.0 $\mu$m, z-offset=0.85 $\mu$m
(dashed line).

\newpage

\vspace{2cm}

\vspace{2cm}

\vspace{3cm}

\newpage
\vspace{2cm}

\vspace{2cm}

\begin{thebibliography}{99}
\bibitem{Attocube}
Staveley Sensors piezotube is mounted on top of an Attocube 3D
positioner ANPxyz100/LIN/LT/HV equipped with the optical position
redout.
\end{thebibliography}
\end{document}